a

# Digital Emotion Regulation on Social Media

Akriti Verma*, Shama Islam*, Valeh Moghaddam†, and Adnan Anwar†
*School of Engineering, Deakin University, Australia  †School of Information Technology, Deakin University, Australia

*Abstract*—Emotion regulation is the process of consciously altering one's affective state, that is the underlying emotional state such as happiness, confidence, guilt, anger etc. The ability to effectively regulate emotions is necessary for functioning efficiently in everyday life. Today, the pervasiveness of digital technology is being purposefully employed to modify our affective states, a process known as digital emotion regulation. Understanding digital emotion regulation can help support the rise of ethical technology design, development, and deployment. This article presents an overview of digital emotion regulation in social media applications, as well as a synthesis of recent research on emotion regulation interventions for social media. We share our findings from analysing state-of-the-art literature on how different social media applications are utilised at different stages in the process of emotion regulation.

## I. Introduction

ONE-third of music streaming on smartphones is intended to influence affective states (such as uplifting motivation during mundane tasks or to purge negative emotions) [1]. According to a recent survey done by a group of researchers in Spain, 73.8% of respondents reported using online music as a means of managing their emotions [2]. Using social media applications was the most effective technique to self-regulate mood during the COVID-19 pandemic. Pre-pandemic, a different survey found that 87% of teens and adults would go online for ways to deal with stress, anxiety, or depression, but just 20% of them would consider seeking professional help [3]. Some instances of emotions driving smartphone usage in daily life include listening to music while exercising, watching online videos during daily commutes, scrolling through social media while waiting in a queue, and browsing shopping websites at the workplace.

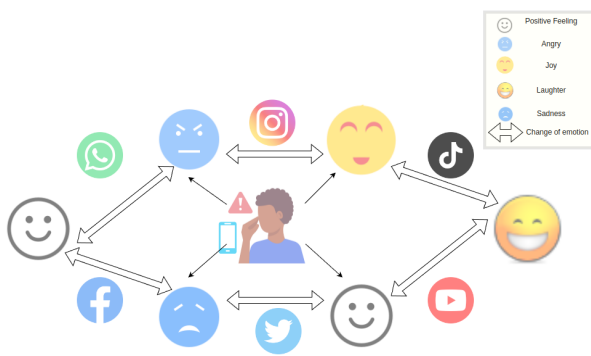

Fig. 1. Digital Emotion Regulation using Social Media

The practice of consciously modifying one's affective state is called emotion regulation (ER). The ability to successfully perform emotion regulation is essential to function effectively in everyday life, to act appropriately in everyday interactions, or merely for hedonic purposes. The ubiquity of digital technology has produced a plethora of prospects for understanding, managing and influencing the world we live in. Digital technology is now being used judiciously to influence our affective states (such as emotions, mood, and stress levels), a process known as digital emotion regulation (DER) In addition to its novelty and scope, understanding digital emotion regulation can facilitate the advancement of the ethical design, development, and application of technology.

The rise in popularity of social media applications has led to significant virtualisation of the day-to-day activities in our lives. Since these applications provide a variety of dimensions for expression and consumption, traditional forms of engagement, leisure and amusement have also been revolutionised by this digitalisation. The COVID-19 pandemic made it difficult to socialise offline, thus more people than ever are choosing to spend their social lives online. Emotions are intertwined in our daily affairs and shape our interactions because they form an essential component of human behaviour influencing how we think and act. Our experiences elicit emotional responses, and these emotions frame our actions. The same is also true for digital technology and social media. It has been observed that emotional states impact interactions with technology, such as joystick movements, typing errors, as well as typing speed. Recent studies have established a correlation between emotional states and the number of application launches, duration of application use, and type of application used, demonstrating that there is a bidirectional relationship between emotions and social media use [4]. People seem to have created a toolkit that allows them to actively appropriate and integrate a variety of applications to manage emotions in everyday life [5]. Scrolling through social media for distraction, watching cat videos for stress relief, texting a friend for support, following an inspirational person to focus on success, and sharing emotions through status updates are a few examples of DER via social media platforms as shown in Figure-1. The omnipresence of social media applications has optimised our ability to regulate emotions at any time or place by allowing us to fine-tune our strategies for specific situations and engage in emotion regulation more frequently.

This emphasises the importance of researching how emotion regulation could be utilised to ensure consistent well-being online by enabling ethical design and development in social media applications. As a result, this work sheds light on recent advances in DER for social media applications. It focuses on the applications of "identifying and understanding" emotion regulation on these platforms and also provides an account of the extrinsic emotion regulation strategies used by users on social media applications. Therefore, the main contributions of this paper can be summarised as follows:

- Providing an overview of the recent research on DER in social media applications, as well as the tools and technological solutions for ER recognition.
- Identifying the knowledge gaps by reviewing a range of literature articles on the various approaches used to study DER in social media and its applications, including the necessity to conceptualise ER on social media and the absence of transfer support.
- Presenting how social media platforms are being utilised as a tool to influence the affective states of others by highlighting the importance of extrinsic emotion regulation.

The remainder of this article is organised as follows. The evolution of DER in social media is discussed in Section-II, followed by a description of the popular approaches used to analyse emotions on social media (Sections III & IV). Then we present some applications of understanding DER in social media (Section-V). Henceforth, we describe how users on social media use extrinsic ER and why it's important for DER (Section-VI). Finally, we identify gaps in the literature and discuss some future directions (Section-VII).

## II. DIGITAL EMOTION REGULATION

Understanding emotion regulation necessitates learning about the emotion regulation process as well as the motivation for doing so. By classifying emotion regulation strategies into five categories according to where they appear in the sequence of emotion generation, Gross's process model describes how emotion regulation is performed. [6]. Situation selection is the first intervention, which entails going into situations that might evoke desired emotions or staying away from scenarios that might induce undesirable emotions. Situation modification makes it possible to control emotions by altering specific aspects of a situation directly after it has been encountered and before an emotional reaction has formed completely. This is accomplished by shifting one's focus from or toward emotion-relevant features deliberately, or by reevaluating a situation to change the way of its emotional impact through cognitive change. Lastly, response modulation can be used to change the physiological, behavioural, or experiential components of an emotional response even after it has already occurred (e.g., exhibiting intensified facial expressions). Figure-2 describes the process of emotion regulation and the strategies a social media user may employ to modify their affective state. Tamir's motive taxonomy, which divides the goal of emotion regulation into two categories, hedonic and instrumental, with additional classifications, may be used to explain emotion regulation motivations [6]. Hedonic motives (Prohedonic and Contrahedonic) entail engaging in emotion regulation in order to experience or avoid certain emotions; typically, pleasant emotions are enhanced while painful emotions are diminished. Instrumental motives (Performance, Epistemic, and Social) are motivations that people have when they think that a particular affective state could help accomplish a task, meet a performance target, or demonstrate appropriate expressions and behaviours [6].

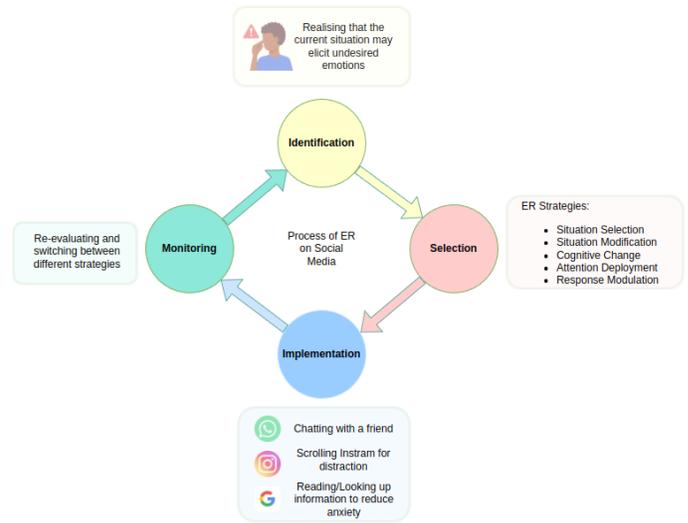

Fig. 2. Process of Emotion Regulation

The examples discussed thus far in this article are cases of intrinsic ER, wherein individuals seek to alter their personal affective states, but, in digital technologies, extrinsic ER is an essential component, where people intentionally try to influence the emotional state of others. Social media platforms allow users to create identifiable profiles, represent public connections and consume, produce, and/or interact with a wide range of user-generated content from their connections on the site. These applications consistently harbour communication among millions of people from all over the world and hence contain traces of emotional expression in abundance. Since the virtual world offers an opportunity to network with people beyond geographical boundaries, exposure to the expressions of others amplifies. As a result, the emotions, actions and online behaviours of the users can be strongly influenced. Given that user engagement is a key factor in the success of social media platforms, and exposure to emotion increases it, social media companies have strategically used the emotional effects of activities on social media platforms to improve user experience and fuel user participation. This intervention by digital media companies to boost the intensity and frequency of expression has resulted into increased online emotion contagion, which is the spontaneous spread and convergence of emotion based on exposure [7], [8]. According to ER psychological research, emotion contagion can be caused by extrinsic emotion regulation [9]. Online movements such as #MeToo and toilet paper hoarding during COVID-19 (#panicbuying), which rely on users' connected behaviours and have an impact on users' offline lives as well, emerged as a result of emotional contagion. Implicit emotion regulation, which does not involve a conscious desire to alter emotional responses, may also play a role in the development of connective action. Nonetheless, this disproportionate induction and convergence of emotion, generates social dysfunction and online toxicity.

There is still a substantial difference between online and offline social behaviour, even though the boundary between the real and the virtual is dissolving as technology becomes an inevitable part of our life. Although online toxicity has been well understood, recognised, and intensively studied for a few years now, remedial measures are challenging to implement due to the nature of online media environments, where the context of user behaviours cannot always be extracted [10], [11], [12]. Anonymity in social media platforms keeps the users behind the safety of a keyboard. This lack of accountability in social media platforms has adversely impacted online well-being. What is acceptable online (due to lack of mediation, moderation, or other restrictions), may not be accepted offline. As a result, it is crucial to understand and encourage responsible use of digital technology to regulate emotions effectively.

## III. EMOTION REGULATION IN SOCIAL MEDIA

Human Computer Interaction (HCI) research has seen an increase in interest in the design and development of technological interventions that can help and support emotion regulation. This renewed interest originates from the realisation and recognition that effective emotion regulation is a vital aspect of mental health. Over the last decade, there has been a lot of research into technology-enabled mental health support and the development of a diverse array of tools designed to assist with emotion regulation [6], [5], [10]. The availability and popularity of low-cost wearable sensors and extensively configurable bio-feedback devices fuelled this. Another factor that has contributed to this advancement is psychological research in the field of emotion regulation. It now recognises effective ER as a key factor in personal well-being and is using it as a trans-diagnostic intervention for mental health disorders. This section summarises the recent elements of research and development that have emerged in the field of digital emotion regulation.

### A. Observational Studies

Recent studies have examined how people combine a variety of social media applications, such as video streaming platforms, discussion forums, online games, and music, to regulate their emotions, thoughts, and behaviours. A common technique known as 'mental reset' has been observed, in which individuals employ social media apps to distract themselves from overwhelming emotions. Several diary studies have been conducted to better understand various aspects of everyday emotion regulation in isolation, such as the use of social media to avert homesickness and university students' use of music streaming platforms. Individuals' multitasking and passive scrolling habits on social media apps have also been explored in studies revealing how people voluntarily take breaks from social media to mitigate negativity or uphold a sense of equilibrium, as well as highlighting the practice of interpersonal emotion regulation in discussion forums [5]. It was discovered that active social media usage had a higher rate of emotion regulation than passive usage, which provided superior support for rest and recovery [13]. The effect of the pandemic on young people's digital emotion regulation habits has also been studied, and it indicates that while the lockdown made people's emotion regulation practises more uniform, it also increased their reliance on technology and increased their proclivity to use emotion suppression strategies [14]. Altogether, these studies show that a variety of digital technologies are employed by people for emotion regulation in everyday life. They emphasise the importance of the technologies packed into these devices and the need to boost well-being online, in addition to the evergrowing role that digital technologies play in fostering mental health globally [10], [6], [5].

### B. Design and Evaluation of Novel Tools

This body of research involves the development of interventions that seek to support, develop, or guide emotion regulation skills or help individuals use such skills in challenging situations. The process of emotion regulation occurs in four stages: recognising the need or realising the desire for emotion regulation, selecting an appropriate strategy, implementing it, and then monitoring the regulated state to determine whether additional regulation is required. Technology-enabled interventions have aimed at either supporting a specific ER strategy or boosting emotional awareness during the identification or monitoring stages [10]. They include experience-based design components which primarily focus on bio-feedback or implicit target responses to nudge users subconsciously toward specific physiological states via haptic interaction, such as imitating heart rate to boost performance by decreasing anxiety [5]. These involve features for both on-the-spot and offline support. Triggers are used to direct or support users during the emotion regulation processes in on-the-spot support systems, which can provide either one-time reminders or ongoing haptic support. Offline support systems use interactivity to create a custom cycle for regulating emotions, such as by visualising a timeline of emotional occurrences to structure users' reflections. Recent advances have also included didactic intervention elements that rely on reminder-based recommender systems, such as suggesting specific ER strategies to users and encouraging them to consider their emotional responses. These works investigate new design possibilities for DER, and by examining how these designs affect users, they offer a new set of directions for this field of study.

### C. Recognising Emotion Regulation

Aside from an individual's desire to regulate their emotions, the ER process involves their surroundings, a situational trigger for emotion, and their attempts to regulate that emotion. It is challenging to measure these variables in a lab setup, so studies have begun investigating ways to recognise them in the wild [1], [5], [2]. To observe the change of emotions using facial expressions, the front camera of smartphones, touch sensors, eye trackers, and motion sensors have been used, both independently and in combination. A recent study used the device's front camera to measure the degree of joy throughout each phone session and divided people into three groups based on how likely they were to feel joy at the start of a session [11]. Go-getters were users who desired a brief sigh of relief or rejoicing before turning on their phone, targeters were users who desired a uniform and gradually increasing joy as the session progressed, and explorers were users who experienced the polymodal joy that sees a gradual decline before locking the phone. Another study applied image manipulation to simulate the realistic image distortion that occurs when capturing a person's expressions for facial expression-based detection of digital emotion regulation [15]. According to research, combining modalities improves the accuracy of affect detection in social media tasks.

## IV. Approaches to Study Digital Emotion Regulation in Social Media

Research in digital emotion regulation requires observing the entire process of emotion regulation, beginning with recognising the need to regulate, evaluating the context, implementing, monitoring, and modifying strategies. The majority of psychological research in ER is based on static self-reported assessments, experience sampling, surveys, or lab-based tasks collected via questionnaires, interviews, and group discussions, making the state-sensitive task of emotion regulation difficult to capture. Participants in this type of data collection are expected to record their interactions with emotion regulation and technology use over a set period of time, and then discuss their use of technology and the emotional reactions that accompany it in an interview. By analysing the expressible, shareable, consumable, and evaluable emotional affordances, it was discovered that, while these emotional affordances yield emotions and responses, they do not play a role in eudaimonic well-being. These methods allow participants to trace and reflect on significant insights by limiting the amount of detail to be recorded but pose a challenge to advance research in the field.

Emotion sensing has gained popularity in HCI research, implying that relatively prevalent technology can be effectively used for emotion recognition. The various sensors (such as accelerometer, gyroscope, proximity sensor, and microphone) built into devices can be used to detect users' internal and external contexts, including emotions. The above tools allow us to better understand how people utilise technology for appropriate emotion-regulation strategies in their everyday life. A growing number of technology devices capable of both passive and active analysis, such as these, are being used to measure specific emotion-regulation processes. Digital technologies, in particular, allow for the comprehensive analysis of instances where individuals engage in regulation strategies and the means they use to do so or adjust their techniques, as well as how well they work in various circumstances and over time.

Methods of exploring DER in social media by combining a smartphone's front camera and touch sensor have also become popular. This data has been used to predict the participants' binary emotional states and can also detect the emotional state while passively using social media. The front camera has been used to measure the degree of joy for a phone session to discover unimodal and polymodal patterns of joy [11]. Researchers contend that by using manipulated images which simulate the realistic image distortion occuring when a person's expressions are captured during phone usage, recognition systems rarely confused expressions of surprise, anger, happiness, and neutrality for other emotions. Attention-based LSTM (Long Short-Term Memory) has also been proposed for a user-independent mobile emotion recognition system that uses smartphone-only or wristband data.

Emotions are inextricably linked to user well-being, according to research, and if these affordances are exploited, our digital well-being is particularly vulnerable. It is alleged that, while social anxiety does not cause problematic social media and smartphone usage patterns on its own, individuals with high levels of social anxiety may prefer online media over face-to-face interactions, which could be the cause of the indirect pathway linking affection-related distorted cognitions and social anxiety.

## V. Applications of Studying Digital Emotion Regulation for Social Media

Because social media forms such a massive part of digital technology, it has inevitably become a part of digital media regulation due to the diverse range of expression opportunities it provides. However, the ease with which emotions can be regulated through digital media does not guarantee their effectiveness, as we see online rumination, trolling, and emotional effusion, which may be caused by emotion dysregulation. Understanding how people use social media platforms to regulate their emotions will help to shape the design and development of strategies to promote effective emotion regulation and improve online well-being. Because of the popularity and widespread use of social media applications, (82.4% of Australians use social media and spend an average of about 2 hours per day online), they can be used to deliver ER-based learning as a takeaway. Today's content moderation is based on identifying specific words or phrases in a post or comment [12]. Understanding how to recognise emotion regulation in online activities will also provide a factual foundation for content moderation.

### A. Enhance Online Well-Being

For the past few years, there has been debate about online toxicity and the negative effects of digital technologies. The idea that emotion can be influenced by technology, that it plays a significant role in the user experience, and that it can be catered through design is a fundamental component of contemporary HCI. Research on user interface design aims to support favourable emotional interactions with digital tools and services, whereas affective computing research aims to detect and adapt to user emotions. Psychology lessons have recently been applied to HCI to develop applications that support and enhance emotional well-being. Digital emotion regulation will not only help understand the relationship between emotions and technology use but will also inform the conceptualization of emotion online, which will eventually enable the creation of services that enrich online well-being [16].

## B. Inform Content Moderation

The vast majority of DER research has been conducted in controlled lab settings, with observation data derived from diary studies, interviews, experience sampling, and remote sensing. These techniques provide insights into the timeline of changes in user experience while also capturing the rich context-dependency and complexity of emotions in their natural state. The sensitive nature of this data, however, makes it impractical for sharing and ineffective for the research community, preventing further development in the area. It has been identified that people use social media as an ER tool. Social media contains a wealth of data containing the emotional trajectories of users worldwide. As a result, exploring ER for social media platforms will help in determining emotion regulation in online activities, which will eventually inform strategies for moderation and control of emotion dysregulation online. Because social media data is easily accessible, it will stimulate contrast and comparison.

## C. Emotion Regulation Transfer Support

The transfer of learning is critical for DER interventions. The impact, sustainability and efficiency of these systems depend on people being able to manage daily situations, without ongoing support. The DER technologies currently include applications that serve as reminders and didactic tools (module-based). Few experiential applications that provide bio-feedback focus on one or more ER strategies (usually suppression, which is not considered ideal for long-term use), which makes the support for ER strategies inconsistent. Understanding and recognising ER in social media activities will make it easier to determine the emotion regulation intervention goals that systems intended to support ER may aim to impact, as well as how/when the interventions can play a role in the ER process [10]. The 'how' here refers to the mode of delivery, which could be didactic or experiential, whereas the 'when' refers to whether the delivery is done offline or on the spot. Examining how people regulate their emotions on social media will empower researchers in developing novel ways to reinforce theory-driven ER targets using novel intervention delivery methods.

## VI. SIGNIFICANCE OF EXTRINSIC EMOTION REGULATION

Owing to the opportunities afforded by the virtual environment as well as the mechanisms by which it is sustained, extrinsic ER is a significant component of digital emotion regulation. Exposure to emotion has been shown to keep users engaged in social media platforms. According to Facebook, the 'withdrawal effect' occurs when users become less engaged on social media as a result of experiencing fewer emotions online. To encourage their users to express emotions, social media platforms incentivise emotional exposure through positive reinforcement such as likes or shares. Nowadays, a substantial part of social media is about influencing others. Figure-3 describes some examples of how social media platforms are being used to influence others. Social media applications are used by both individuals and companies as platforms for profitable marketing strategies. User-Generated Content is at the heart of many of today's most popular online social networks, including Instagram, YouTube, Twitter, and Twitch. The success of social media influencers is determined by how well they can affect or influence their audience and how long they can do so. Given the extensive exposure to other people's emotions via digital media, it appears that the contagious transmission of digital emotions is having a significant impact on user emotions and behaviour.

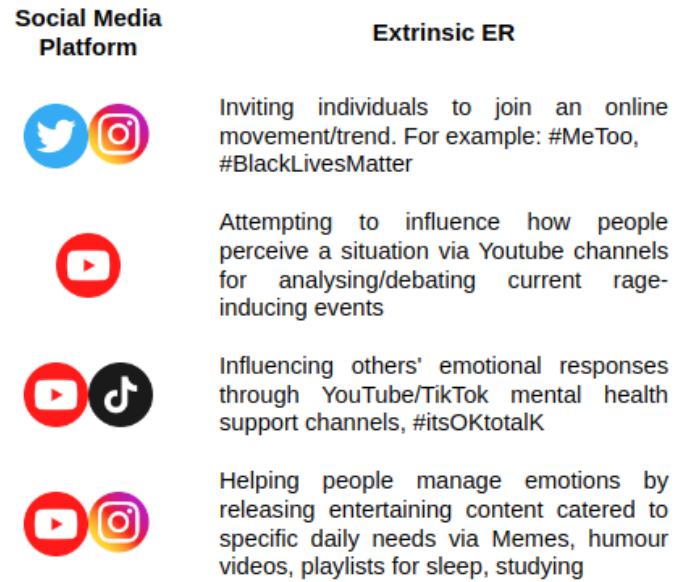

Fig. 3. Extrinsic ER on Social Media

Online movements that rely on users' connected behaviours and have an impact on offline life, such as #MeToo or hoarding toilet paper during COVID-19, have emerged as a result of the increased emotional contagion online. Emotion contagion, as described earlier in this article, maybe the outcome of extrinsic ER, a phenomenon in which an individual or a group of individuals may purposefully seek to influence the emotions of others in order to achieve their goals. Although these incidents can be regarded as simply empathising or relating to the emotions being expressed, the fact that the original trigger is usually from a few, and in some cases, a single user, and that there is an intention of influencing the audience, confirms the likelihood of extrinsic DER. These events cause a disproportionate induction and convergence of emotion, which results in social dysfunction and online toxicity.

According to studies, toxicity online is not concentrated among a few highly toxic users, but rather among a large number of moderately toxic ones. When exploring how people use technology for emotion regulation in everyday life, it was discovered that, in addition to turning to digital media after realising the need for ER, users can also experience undesirable emotions online and may use the convenience of virtual platforms to undergo ER. A digitally induced emotional challenge, for example, can be most simply dealt with by a digital response, such as quelling anger on a post [5]. Increased hate speech and internet trolling are two further manifestations of the same trend. Online toxicity has been widely acknowledged, recognised, and extensively researched for a few years. However due to the nature of online media platforms, where the context of user behaviour cannot always be extracted, mitigation measures are challenging to put into practise. To ensure internet users' well-being, it is critical to determine when and how effective extrinsic ER or the resulting emotional contagion might become detrimental to online health.

## VII. Conclusion and Future Work

Since emotion regulation is such an essential component of overall health, it is necessary to develop interventions that can help people shift their dominant (or modified) coping mechanisms to digital forms in ways that are broadly applicable and manualized. The desire to regulate one's emotions has been linked to the problematic use of social media and smartphones. Users spend hours scrolling through social media sites to relieve stress (engaging in unhealthy strategies for emotion regulation like rumination, overthinking or self-blame), but they end up feeling distressed because of the content they consume and the amount of time they waste procrastinating. A lack of emotion regulation skills may be the root cause of problematic internet use. As a result, we believe the following potential future research directions in digital emotion regulation will help bridge the gap between emotions and technology usage.

### A. Developing Digital Solutions

Current emotion regulation tools include mood-based recommendation systems and reminders, which are only temporary and difficult to integrate into daily life. Furthermore, studies show that a combination of online media platforms and apps are used to perform digital emotion regulation. However, these apps or platforms have not been curated for the purpose. Innovative digital solutions that facilitate the transmission and development of emotion regulation practises will not only have a long-term impact, but will also boost social media usage. It is necessary to develop sophisticated tools that allow people to consider, evaluate, and self-diagnose the positive and negative outcomes of their digital activities. There is a lack of techniques for interaction design that use existing or new mechanisms to help users navigate emotional experience trajectories rather than providing recommendations based on their emotional state.

### B. Synthesising Data and Prototypes for Contrast and Comparison

The majority of research in the field of digital emotion regulation is based on field studies, ecological momentary assessments (EMA), or physiological sensors combined with facial data, as discussed earlier in the article. Due to their sensitive nature, these datasets cannot be shared, resulting in a lack of data to compare and contrast the results produced, as well as a lack of a common prototype for synthesising emotion regulation. The use of easily accessible data for studying DER will facilitate contrast and comparison.

### C. Recognising DER

Although the relevance of situation and context in the process of regulating digital emotions is widely acknowledged, little research in this area has been conducted through field studies and interviews. Understanding a user's online presence and connecting it to their approach to emotion regulation would provide new insights into how technology can help break down and personalise existing cognitive modules into manageable pieces. Empirical studies of people's usage of digital technologies to achieve desired emotional results are required to learn how users can be supported by digital media when undergoing emotion regulation. Because of the complexity of emotional expressions, a large portion of DER research presents qualitative results that are difficult to extend. This opens up opportunities for the use of machine learning and deep learning-based algorithms, which can be used to understand the habits, frequency, and intensity of DER practises.

Researchers have addressed concerns about the negative effects of excessive technology use on mental health. Although knowledge of digital ER may not eliminate these concerns, it does imply that some digital media may be used in conjunction with constructive psychological techniques. Learning more about the regulation of emotions digitally can help resolve debates about the excessive use of technology by emphasising that people's motivations for using technology. Seemingly counterproductive activities, such as distraction, may not always be hedonistic but instead serve significant instrumental purposes, such as improving work performance or fostering social harmony.